\documentclass[10pt,pra,twocolumn,notitlepage,showpacs,superscriptaddress]{revtex4-1}
\usepackage{amsthm,amsmath,amsfonts,amssymb,verbatim, color}
\usepackage{graphicx}
\usepackage{bm}
\usepackage{epsfig}
\usepackage[colorlinks=true,citecolor=blue,linkcolor=blue,urlcolor=blue]{hyperref}
\begin{document}

\title{Electronic cooling in Weyl and Dirac semimetals}
\author{Rex Lundgren}
\email{rexlund@physics.utexas.edu}
\author{Gregory A. Fiete}
\affiliation{Department of Physics, The University of Texas at Austin, Austin, TX 78712, USA}
\begin{abstract}
Energy transfer from electrons to phonons is an important consideration in any Weyl or Dirac semimetal based application. In this work, we analytically calculate the cooling power of acoustic phonons, i.e. the energy relaxation rate of electrons which are interacting with acoustic phonons, for Weyl and Dirac semimetals in a variety of different situations. For cold Weyl or Dirac semimetals with the Fermi energy at the nodal points, we find the electronic temperature, $T_e$, decays in time as a power law. In the heavily doped regime, $T_e$ decays linearly in time far away from equilibrium. In a heavily doped system with short-range disorder we  predict the cooling power of acoustic phonons is drastically increased because of an enhanced energy transfer between electrons and phonons. When an external magnetic field is applied to an undoped system, the cooling power is linear in magnetic field strength and $T_e$ has square root decay in time, independent of magnetic field strength over a range of values.
\end{abstract}


\maketitle

\section{Introduction}
Dirac \cite{2007NJPh....9..356M,PhysRevLett.108.140405} and Weyl \cite{PhysRevB.83.205101} semimetals have received an enormous amount of attention due to the wide range of exotic physical phenomena they are theoretically predicted to host. For example, exotic edge states known as Fermi arcs \cite{PhysRevB.83.205101,PhysRevB.84.245415,PhysRevB.86.195102,PhysRevB.87.245112,2014arXiv1401.0529H,PhysRevB.89.235315,2015arXiv150104998Z} have recently been experimentally seen on the surface the Dirac semimetals $\mathrm{Na}_3\mathrm{Bi}$ \cite{Xu16012015} and $\mathrm{Cd}_3\mathrm{As}_2$ \cite{2014NatSR...4E6106Y} after their theoretical prediction from first-principles calculations \cite{PhysRevB.88.125427,PhysRevB.85.195320}. There also is recent experimental evidence of a Weyl semimetal phase in TaAs \cite{2015arXiv150203807X,2015arXiv150204684L,2015arXiv150200251Z} and photonic crystals \cite{Lu07082015}, after their theoretical predictions \cite{PhysRevX.5.011029,2015arXiv150100755H,2013NaPho...7..294L}. Weyl (Dirac) semimetals have linearly dispersing excitations [which obey the Weyl (Dirac) equation, respectively] around the band touching points referred to as Weyl (Dirac) nodes. These nodes possess non-zero Berry curvature \cite{Volovik}, which gives rise to nontrival momentum-space topology. Weyl semimetals also have many interesting topological properties, including the chiral magnetic effect \cite{PhysRevD.78.074033,2014PrPNP..75..133K} and other phenomena associated with the chiral anomaly \cite{PhysRev.177.2426,1969NCimA..60...47B}. The chiral magnetic effect is the separation of electric charge along the direction of an applied external magnetic field and occurs when band touching points have different energies. There is recent experimental evidence for the existence of the chiral magnetic effect in $\mathrm{ZrTe}_5$ \cite{2014arXiv1412.6543L}. The chiral anomaly causes the number of particles with a given chirality to not be conserved and occurs when external parallel electric and magnetic fields are applied. Dirac semimetals can be topologically protected by space group symmetries \cite{PhysRevLett.108.140405}, however they are generally not as stable as Weyl semimetals. For an overview of Weyl and Dirac semimetals, see Refs.~\cite{2013CRPhy..14..857H} and \cite{2013arXiv1301.0330T}.

We are interested in exploring energy exchange between electrons and phonons in Weyl and Dirac semimetals when the electrons and lattice are at different temperatures. Energy exchange with phonons is often the most dominate energy exchange mechanism in condensed matter systems \cite{PhysRevLett.59.1460}. As a result, energy transfer from electrons to phonons is a key issue with possible Weyl and Dirac semimetals based devices that take advantage of the topological properties or Berry curvature of Weyl and Dirac semimetals. Typically, to probe energy relaxation, electrons are excited to high temperatures using an optical laser pulse. The hot electrons will then equilibrate through electron-phonon interactions and the electronic temperature will approach the temperature of the lattice. As in normal metals and graphene, we assume electron-electron interactions rapidly thermalize the electrons among themselves during the relaxation process \cite{Mahan,PhysRevLett.101.157402,:/content/aip/journal/apl/92/4/10.1063/1.2837539}. Transport measurements also provide a way to study electron-phonon interactions in Weyl semimetals, but as with graphene \cite{PhysRevLett.100.016602}, resistivity due to electron-phonon scattering is expected to be smaller than the residual resistivity contribution that arises from disorder or electron-electron interactions \cite{PhysRevB.90.165115,2014arXiv1412.2767S}.

In this paper, we analytically study the energy transfer of electrons to acoustic phonons in Weyl and Dirac semimetals in a variety of situations. While we focus on acoustic phonons, we note that optical phonons will play a dominant role in cooling for electronic temperatures around and above the lowest optical branch. First principles calculations predict the optical branch to have a frequency of around 3.5 THz for BLi, a material that is expected to host a Weyl semimetal phase \cite{2014arXiv1412.2607D}. Assuming other Weyl/Dirac materials have a similar scale, our results should be applicable below some temperature range on the order of a few hundred Kelvin. The precise temperature range at which acoustic phonons dominate over optical phonons depends on the electron density and lattice and electron temperatures along with electron-phonon coupling strength.  We are unaware of any experimental data on electron-optical phonon coupling strength for Dirac/Weyl materials, so we leave a discussion on the competition between cooling power of acoustic and optical phonons for future work. We stress, however, there should be a temperature (which is below the temperature of the lowest optical branch), at which acoustic phonons dominate cooling and our results apply. 

Alternatively, one could use our results to investigate the cross over between acoustic and optical phonon dominated cooling. By first experimentally observing the results mentioned here for very low temperatures, one could then increase the electronic temperature until a change in the cooling properties is observed. When the chemical potential is at the nodal point, we find the temperature of the electrons decays as a power law in time over a few microseconds for $\mathrm{Cd}_3\mathrm{As}_2$, for example. These long-lived hot carriers (compared to a characteristic time scale of picoseconds in metals \cite{PhysRevLett.58.1680} when $T_e$ is greater than the Debye temperature, $T_D$)--important in calorimetry and bolometry \cite{RevModPhys.78.217}--exists as long as $T_e$ is less then the temperature of the optical branch, whereas in normal metals long lived hot carriers only exist for very low temperatures ($T_e\ll T_D$) \cite{Mahan,PhysRevLett.55.422}. 

In the highly doped limit, we find the temperature of the electrons decays linearly in time far from equilibrium and exponentially decays in time near equilibrium.   Motivated by recent electron cooling experiments on the two-dimensional analog (in some respects) of Dirac and Weyl semimetals, graphene, we also consider the effect of short-range disorder in the heavily doped regime. For graphene, short-range disorder greatly increased the cooling power due to enhanced energy transfer between electrons and phonons \cite{PhysRevLett.111.027403,2013NatPh...9..109B} and the relaxation rate can be controlled by varying disorder \cite{PhysRevLett.109.106602}. We show that such an enhancement of cooling power should be expected over a wide range of temperatures in $\mathrm{Cd}_3\mathrm{As}_2$ and other related materials. This result allows for a new three-dimensional material whose cooling properties can be controlled by disorder.  Finally, in the presence of a moderate strength external magnetic field, the power loss of electrons is found to be linear in magnetic field strength and the temperature of electrons linearly decays as a square root in time with a rate independent of the external magnetic field.

Our paper is organized as follows. In Sec.~\ref{sec:single_node}, we calculate the power loss due to acoustic phonons for a single Weyl node. In Sec.~\ref{sec:disorder}, we consider the effect of short-range disorder on electronic cooling. In Sec.~\ref{sec:mag_field}, electronic cooling due to acoustic phonons is investigated in the presence of an external magnetic field. Finally in Sec.~\ref{sec:CON}, we present our conclusions. Some technical results are regulated to the appendices.
\section{Single Weyl node}\label{sec:single_node}
We first consider the case of a single isotropic Weyl node. To generalize to $N$ Weyl nodes, one can multiply the result for a single Weyl node by $N$ (provided there is negligible scattering between nodes). To obtain the results for a single Dirac node, multiply the cooling power for a single Weyl node by two. Our approach follows the one taken in Ref.~\cite{PhysRevLett.59.1460} for normal metals and Refs.~\cite{PhysRevLett.102.206410} and \cite{PhysRevB.81.245404} for graphene. The power loss, $P$, is given by 
\begin{equation}
P=\frac{\partial E}{\partial t}=\partial_t\sum_{\vec{k},\alpha}\epsilon_{\vec{k},\alpha}f^{\alpha}_{\vec{k}},
\label{General_P_Loss}
\end{equation}
where $f_{\vec{k}}^{\alpha}$ is the time-dependent Fermi distribution function, $E$ is the energy of the system, $\epsilon_{k,\alpha}=\alpha\hbar v_F|\vec{k}|$ is the quasiparticle dispersion relation for quasiparticles with wavevector $\vec{k}$, $v_F$ is the Fermi velocity, $\hbar$ is the reduced Planck's constant, and $\alpha=\pm 1$ labels the valence and conduction bands. The Fermi velocity has been experimentally found (via transport, optical measurements, angle-resolved photoemission spectroscopy) and theoretically predicted (via first principal calculations) to range from $1\times10^5~m/s$ to $2\times10^6~m/s$ \cite{PhysRevB.87.235121,Liu21022014,2014NatMa..13..851J,2014NatCo...5E3786N,PhysRevLett.113.246402,2014arXiv1412.0824C,2014arXiv1412.4380L,2014arXiv1412.2607D, 2015arXiv150101175P} in various Dirac and Weyl semimetal systems. In the case of $\mathrm{Cd}_3\mathrm{As}_2$, the Dirac cone is anisotropic \cite{2014NatCo...5E3786N}. We do not expect anisotropy to significantly alter our predictions. 

In this work, we take $v_F=1\times10^6 m/s$, a value appropriate for $\mathrm{Cd}_3\mathrm{As}_2$. Eq.~\eqref{General_P_Loss} can be rewritten as a differential equation for the electronic temperature, $\partial_t T_e=\frac{P}{C_e}$, where $C_e=\partial_{T_e}E$ is the electronic heat capacity. We note that both cooling power and electronic heat capacity scale with the number of Weyl nodes, thus the temporal evolution of $T_e$ will be independent of the number of Weyl nodes, under the assumption of negligible inter-node scattering. From Boltzmann's equation, we have 
\begin{equation}
\partial_tf^{\alpha}_{\vec{k}}=-\sum_{\vec{p},\beta}\left(f_{\vec{k}}^{\alpha}(1-f_{\vec{p}}^{\beta})W_{\vec{k}\alpha\rightarrow\vec{p}\beta}-\{\vec{k}\alpha\leftrightarrow\vec{p}\beta\}\right),
\label{powerloss_NOB}
\end{equation}
where
\begin{align}
W_{\vec{k}\alpha\rightarrow\vec{p}\beta}=\frac{2\pi}{\hbar}\sum_{\vec{q}} [|M_-|^2(N^L(\omega_q)+1)\delta_-+N^L(\omega_q)|M_+|^2\delta_+],
\end{align}
is the transition rate between states $|\vec{k},\alpha\rangle$ and $|\vec{p},\beta\rangle$, $M_{\pm}=w^{\alpha\beta}_q\delta_{\vec{k},\vec{p}\pm\vec{q}}$ is the transition matrix element, $\delta_{\pm}=\delta(\epsilon_{\vec{k},\vec{p}}^{\alpha,\beta}\pm\omega_q)$, $\epsilon_{\vec{k},\vec{p}}^{\alpha,\beta}=\epsilon_{k,\alpha}-\epsilon_{p,\beta}$, $N^{L}(\omega_q)$ is the Bose distribution function evaluated at the temperature of the lattice $T_{L}$, $w^{\alpha\beta}_q=\frac{\hbar^2 D^2q^2(1+s_{\alpha\beta}\cos\theta)}{4\rho V\omega_q}$ \cite{ManyBody,PhysRevLett.102.206410,PhysRevLett.110.046402}, $\omega_q=\hbar c_s q$ is the dispersion relation for the phonons, $c_s$ is the speed of sound, $\theta$ is the angle between $\vec{k}$ and $\vec{p}$, $\rho$ is the mass density of ions, $V$ is the volume, $D$ is the deformation potential constant, and $s_{\alpha\beta}=1$ for intraband transitions and $-1$ for interband transitions. The deformation potential constant is just the electron-ion potential at zero wavevector \cite{Mahan}. 

In this paper, we take $c_s=2.3\times10^3~m/s$ and $\rho=7\times10^3~kg/m^3$ which are the speed of sound and density of $\mathrm{Cd}_3\mathrm{As}_2$ \cite{wang2007computation} unless otherwise noted. The deformation potential has been estimated in $\mathrm{Cd}_3\mathrm{As}_2$ to be in the $10-30$ eV range from transport measurements \cite{PhysRevB.18.4542}. Throughout this work, we take the deformation potential to be $20$ eV. We ignore vertex corrections, which give corrections that scale as $c_s/v_F$ \cite{Mahan,PhysRevB.89.165119}, a small value in realistic systems. After some algebra (see App.~\ref{sec:Weyl_Node} for details), we find the power loss, to lowest order in $c_s/v_F$ and for arbitrary chemical potential, $\mu$, referenced from the nodal point is
\begin{align}
P(\mu,T_e,T_{L})\approx-\frac{VD^2}{\rho }\frac{(k_BT_e)^6}{\pi^3\hbar^7v_F^8}(k_BT_{e}-k_BT_L)\nonumber \\ \times
\int_0^{\infty}\mathrm{d}xx^5\bigg(f(x-\beta_e\mu)+f(x+\beta_e\mu)\bigg),
\label{Gen_Power}
\end{align}
where $k_B$ is Boltzmann's constant. We now discuss some limits of Eq.~\eqref{Gen_Power}. We note, for a general chemical potential, to find the relaxation rate one must solve coupled differential equations (Eq.~\eqref{General_P_Loss} and $\frac{\partial n}{\partial t}=0$, where $n$ is the electronic density which is assumed to be spatially uniform) to find the relaxation rate since the chemical potential has a temperature dependence.

In the limit of $\mu=0$, we have
\begin{equation}
P=-\frac{2V D^2(k_B T_e)^6(k_BT_{L}-k_BT_e)}{\rho \pi^2 v_F^8\hbar^7} \Gamma(6)\eta(6),
\end{equation}
which gives (using $C_e=\frac{4Vk_B^4T^3\Gamma(4)\eta(4)}{\pi^2(\hbar v_F)^3}$, where $\Gamma$ is the gamma function and $\eta$ is the Riemann zeta function),
\begin{equation}
\frac{\partial T_e}{\partial t}=-\gamma_r T_e^3(T_{e}-T_L),~~~~~~\gamma_r=\frac{D^2k_B^3}{2\rho\pi^4 v_F^5\hbar^4}\frac{\Gamma(6)\eta(6)}{\Gamma(4)\eta(4)}.
\end{equation}
We remind the reader that the temperature difference between lattice temperature and electron temperature is due to the system being hit with an optical pulse. This result agrees with the dimensional analysis of $P$, $E$, and Eq.~\eqref{General_P_Loss} put forth in Ref. \cite{PhysRevLett.102.206410}. The cooling power at low temperatures is weak due to the high exponent of $T_e$ that appears in the cooling power. Physically, the weak cooling power of acoustic phonons in Weyl and Dirac semimetals is due to the small energy of acoustic phonons, $\frac{c_s}{v_F}k_BT_e$, at a typical transition momentum of $\frac{k_BT_e}{\hbar v_F}$ and the small density of states for electronic transitions. 
Far from equilibrium, i.e, in the limit that $T_{L}\ll T_e$, we find $T_e(t)=\frac{T_0}{(1+\frac{t}{\tau_0})^\frac{1}{3}}$, where $\tau_0=\frac{1}{3\gamma_r T_0^3}$ and $T_0$ is the initial temperature of the electrons. Taking an initial electron temperature of 140 Kelvin, we find $\tau_0=36\times10^{-6}$ s. 

In the limit where $T_{e}\gtrsim T_{L}$, the electronic temperature decays exponentially with a characteristic time scale, $\tau_L=\frac{1}{\gamma_r T_L^3}$. This should be compared to the low electronic temperature ($T_e\ll T_D$) cooling in metals. In this case, $P\propto T_e^5$ and we have similar slow cooling of the electronic temperature \cite{Mahan}. However, this slow cooling only happens in metals when $T_e\ll T_D$. In contrast, slow electronic cooling in Weyl and Dirac semimetals exist for a wide range temperatures (as long as $T_e$ is much less than the temperature of the optical phonon branch, which is typically on the order of a few hundred Kelvin). We note that Cd$_3$As$_2$ intrinsically has a large amount of charge carriers \cite{2014arXiv1412.4380L,2015arXiv150202264P}. Thus it is an open experimental question if this limit can be reached in Cd$_3$As$_2$.

We now discuss cooling when the system is heavily doped. For $\mathrm{Cd}_3\mathrm{As}_2$, an experimentally well-established Dirac semimetal, has a Fermi energy of around $200$ meV (in the heavily doped limit the chemical potential is the Fermi energy). Unfortunately, this energy scale is nearly the same as same the band inversion energy scale, which is about $250$ meV \cite{PhysRevB.88.125427}, thus it is questionable whether the Dirac fermion description is applicable at finite temperatures for $E_F=200$ meV \footnote{ One may also worry about phonons coupling the electronic bands near the band inversion energy. However, given that angle-resolved photoemisson spectroscopy shows a well defined energy band below $200$ meV in $\mathrm{Cd}_3\mathrm{As}_2$ \cite{2014NatCo...5E3786N}, we believe this concern is not warranted.}. However, recent experimental progress has been made in tuning the Fermi level \cite{2014arXiv1412.4380L,liu2014stable} in $\mathrm{Cd}_3\mathrm{As}_2$ and thus we believe our predictions can be experimentally realized by lowering the Fermi level. To this end, we choose $\mu=100$ meV, a value well below band inversion energy scale. When $k_BT \ll\mu$, the maximum phonon momentum is $2\hbar k_F$, where $k_F$ is the Fermi momentum. Thus, the maximum phonon energy is given by $\hbar c_s k_F$. When the lattice temperature is below $T_{BG}=\frac{\hbar c_s k_F}{k_B}$, the Block-Gr\"{u}neisen temperature \cite{9781139644075}, our approach breaks down. In this respect, the heavily doped case resembles the typical metallic case, where the quasielastic approximation fails below $T_{BG}$ \cite{9781139644075,PhysRevLett.102.206410}. Using $100$ meV for the chemical potential gives a Block-Gr\"{u}neisen temperature of around $5$ K.  Given the recent experimental progress in tuning the Fermi level in $\mathrm{Cd}_3\mathrm{As}_2$, the Block-Gr\"{u}neisen temperature is also tunable parameter. When $k_B T \ll \mu$, we can use the Sommerfeld expansion to evaluate the integral in Eq.~\eqref{Gen_Power}. The cooling power is found to be
\begin{equation}
P\approx-\frac{VD^2\mu^6k_B}{3\rho\pi^3\hbar^7v_F^8}(T_{e}-T_L).
\label{NORMAL_PLOSS_DOPED}
\end{equation}
We also obtain (using $C_e=\frac{Vk_B^2T\mu^2}{(\hbar v_F)^3}$)
\begin{equation}
\frac{\partial T_e}{\partial t}=-\gamma_p\frac{T_e-T_L}{T_e},~~~~~~\gamma_p=\frac{D^2\mu^4}{3k_B\hbar^4v_f^5\rho}.
\end{equation}
When $T_L\ll T_e$, the electronic temperature decays linearly in time with a rate given by $\gamma_p$. Using the experimental values for $\mathrm{Cd}_3\mathrm{As}_2$ we find $\gamma_p=1.8\times10^{10}$ K/s. Thus, the equilibration process is much faster for heavily doped systems compared to undoped systems. Closer to equilibrium, $T_e$ decays exponentially with a rate given by $\gamma_p/T_L$.

\section{ Short-Range Disorder}\label{sec:disorder}
We now consider the effects of short-range disorder on cooling for heavily doped Dirac or Weyl semimetals. Recall in the absence of disorder, the momentum of the phonons are limited to $2\hbar k_F$, and thus the phonons have small energies, $\frac{c_s}{v_F}\mu$. With disorder, phonon momentum is no longer restricted and may reach up to $k_B T/c_s$ \cite{PhysRevLett.109.106602}. This provides a boost to cooling power since the phonons can take away more energy from the electrons. For low impurity concentrations, this process can be described by dressing the electron-phonon vertex. Following the formalism developed in Ref.~\cite{PhysRevLett.109.106602}, we derive the transition matrix elements and analytically find the power loss (see App. \ref{sec:DIS_APP} for a derivation of the transition matrix elements) for disorder described by the following zero-range potential $V(r)=u\sum_{j}\delta(r-r_j)(1+\sigma_z)/2$, where $r_j$ is the location of the $j$th impurity and $\sigma_z$ is the third Pauli matrix. This formalism has had success in describing electronic cooling in graphene.  

As first mentioned in Ref.~\cite{PhysRevLett.109.106602}, allowing for this type of generalized disorder that depends on the spinor structure of the wave-function allows one to obtain a matrix element linear in $u$. In other words, pure scalar disorder gives a quadratic dependence on the disorder strength and thus the power will scale as $u^4$, which is small for weak disorder and won't provide efficient cooling. Physically, this type of disorder might arise from differences of sublattice potential \cite{PhysRevLett.109.106602} or magnetic impurities \cite{PhysRevB.87.155123,1367-2630-15-12-123019}. The transition matrix elements are $|M_{\pm}|^2=|M|^2=\frac{\pi u^2 D^2n_0}{4\rho \hbar c_s v_F^2 q^3}|\langle k'|(\vec{\sigma}\times \vec{q})_z|k\rangle|^2,$ which we plug into Eq.~\eqref{General_P_Loss} and then perform an angular average to find
\begin{equation}
P_d=\frac{V\pi \nu(\mu)D^2k_B^4}{\rho (\hbar c_s)^3v_F l}\frac{\pi^4}{30}(T_e^4-T_{L}^4),
\label{PLOSS_DISORDER}
\end{equation}
where $l$ is the mean free path and $\nu(\mu)$ is the density of states at the Fermi energy. The ratio of power loss for disorder to the normal momentum conserving process [Eq.~\eqref{NORMAL_PLOSS_DOPED}], after linearizing Eq.~\eqref{PLOSS_DISORDER} in $T_e-T_{L}$ is
\begin{equation}
\frac{P_d}{P}=\frac{\pi^6}{120}\frac{1}{k_Fl}\left(\frac{T_e}{T_{BG}}\right)^3.
\label{ratio_power}
\end{equation}
The mean-free path in $\mathrm{Cd}_3\mathrm{As}_2$ is on the order of a $100$ nm \cite{2014NatCo...5E5161P}. We thus take $k_Fl=40$. Disorder assisted cooling then dominates if $T_e\gtrsim 2 T_{BG}$. This result is insensitive to the precise value of $k_Fl$ because of the cubic root in Eq.~\eqref{ratio_power}.  The ratio for arbitrary values of electron and lattice temperature is
\begin{equation}
\frac{P_d}{P}=\frac{\pi^6}{120k_Fl}\frac{T_e^3+T_e^2T_L+T_eT_L^2+T_L^3}{T_{BG}^3}.
\end{equation}
Taking $T_e=50$ K, $T_L=10$ K and $T_{BG}=5$ K, we find that cooling power is enhanced by a factor of 250 in the presence of short-range disorder. The time evolution for $T_e\gg T_L$ is 
\begin{equation}
T_e(t)=\frac{T_0}{\sqrt{1+2 \sigma t T_0}},
\end{equation}
where $\sigma=\frac{\pi^3D^2k_B^2k_F}{60\rho (\hbar c_s)^3 v_F (k_Fl)}$. The cooling time can be controlled by tuning the amount of disorder. This possibility was first suggested in Ref.~\cite{PhysRevLett.109.106602} for graphene. More specifically, the ratio of time it takes to cool to some temperature (which is greater than the lattice temperature) for two different disorder strengths is the ratio of the mean free paths.

We note that scattering from Coulomb impurities will likely dominate electronic transport due to inefficient screening of three dimensional semimetals \cite{PhysRevB.90.165115,2015arXiv150103815R}, however we expect electronic cooling by acoustic phonons to be dominated by short-range disorder as in graphene \cite{2014arXiv1410.5426S}. Physically, this is due the fact that Coulomb disorder doesn't have any spinor structure and vanishes to first order in disorder strength.

\section{ Electron cooling in a magnetic field}\label{sec:mag_field}
In this section we consider the effect of an applied external magnetic field on electron cooling in the cold, neutral limit. Most of the interesting physics of topological semimetals involve the presence of an external magnetic field \cite{PhysRev.177.2426,1969NCimA..60...47B,PhysRevB.89.245103}. We consider the following low-energy Hamiltonian of a Weyl semimetal with two nodes in an external magnetic field (with $\hbar=1$) \cite{PhysRevB.89.085126},
\begin{equation}
H=\int\mathrm{d}^3r \bar{\psi}\left(-iv_F(\gamma\cdot(\nabla+ie\vec{A})-(\vec{b}\cdot\bold{\gamma})\gamma^5\right)\psi,
\end{equation}
where $\vec{A}$ is the vector potential, $\psi$ and $\bar{\psi}=\psi^{\dagger}\gamma_0$ are four component spinors, $e$ is the electric charge, $\gamma$ are the four-dimensional Dirac matrices in the chiral representation and $\gamma^5=i\gamma^0\gamma^1\gamma^2\gamma^3$.  The Weyl nodes are separated in momentum space by the vector $2\vec{b}$. We take the two Weyl nodes to be at the same energy, i.e. the zero-component of the four vector $\vec{b}$ is zero. Our results also apply for Dirac semimetals when $b=0$. Recall Weyl nodes at different energies give rise to the chiral magnetic effect, which generates an electrical current parallel to the external magnetic field. This electrical current will cause phonon drag \cite{Gurevich1989327}. Thus, the phonons will dissipate energy along with transporting it. While this situation is very interesting, it is beyond the scope of this work. The Weyl nodes are taken to be separated in the $z$-direction and the magnetic field, $\vec{B}$, is also taken to lie in the $z$-direction. We leave any possible dependence of the cooling power on the angle between $\vec{b}$ and $\vec{B}$ for future work. We ignore the effects of the magnetic field on the ions due to their large mass.

The cooling power can be written in terms of the imaginary part of the self-energy as originally derived by Kogan \cite{Kogan}. The power loss is given by
\begin{equation}
P=\sum_{\vec{q}}\int\frac{\mathrm{d}\omega}{\pi}\omega (N^L(\omega)-N^e(\omega))\mathrm{Im}\Pi^{\mathrm{Ph}}(\vec{q},\beta_e,\omega)\mathrm{Im}D(\vec{q},\omega),
\label{KOGAN}
\end{equation}
where $\mathrm{Im} \Pi^{\mathrm{Ph}}(\vec{q},\beta_e,\omega)$ is the imaginary part of the phonon self-energy and $\mathrm{Im}~D(\vec{q},\omega)=\pi(\delta(\omega-\omega_q)-\delta(\omega+\omega_q))$ is the imaginary part of the phonon Green's function \cite{PhysRevB.41.3561}. The phonon self-energy to one loop order is given by \cite{Mahan},
\begin{eqnarray}
\Pi^{\mathrm{Ph}}(\vec{q},i\omega_n)=\hspace{6cm}\nonumber \\
\frac{g^2(\vec{q})}{\beta_e V}\sum_{ip_m}\sum_{\vec{k}}\mathrm{Tr}[\gamma^0G(\vec{k},ip_m)\gamma^0G(\vec{k}+\vec{q},i p_m+i\omega_n)],\nonumber \\
\end{eqnarray}
where $G(\vec{k},ip_m)$ is the Greens function of the electrons in the presence of an external magnetic field, $g^2(\vec{q})=\frac{D^2q}{4\rho c_s}$ is the electron-phonon coupling strength and $\beta_e$ is the inverse electron temperature. We derive the imaginary part of the phonon self-energy in the presence of a magnetic field in App.~\ref{sec:MAG_APP}. We find the exact expression for the imaginary part of the lowest Landau level contribution to the self-energy (after analytically continuing to real frequencies) is,
\begin{align}
\mathrm{Im}^{\mathrm{Ph}}[\Pi(\vec{q},\Omega)]=\sum_{\lambda=\pm}\frac{g^2(\vec{q})\Omega}{8\pi^5v_Fl_B^2}\delta(-\Omega-\lambda v_Fq_z)e^{-\frac{q_{\perp}^2l_B^2}{2}},
\label{Img_Self}
\end{align}
where $l_B=\sqrt{\frac{1}{eB}}$ is the magnetic length and $q_\perp^2=q_x^2+q_y^2$. We observe that the imaginary part of the self energy is independent of temperature and chemical potential. A similar feature is seen in the current-current correlation function for Weyl semimetals in the presence of an external magnetic field \cite{PhysRevB.89.085126}.

After plugging Eq.~\eqref{Img_Self} into Eq.~\eqref{KOGAN}, the contribution of the lowest Landau level at $\mu=0$ at a lattice temperature of zero to the power loss is (restoring factors of $\hbar$)
\begin{equation}
P_B\approx-\frac{VD^2\omega_D^5}{320\pi^7v_F^2l_B^2\rho c^4},
\end{equation}
where $\omega_D$ is the Debye frequency (See App.~\ref{sec:MAG_APP} for some keys steps in this derivations). In this case, $\omega_D$ acts as a high-energy cut off. This result is valid to lowest order in $\frac{c_s}{v_F}$ and in the limit $\alpha_B^2\gg\frac{1}{2}\frac{v_F}{c_s}\frac{T_D}{T_e}$, where $\alpha_B=\frac{\hbar v_F \frac{1}{l_B}}{k_BT_e}$. The limit that $\alpha_B^2\gg\frac{1}{2}\frac{v_F}{c_s}\frac{T_D}{T_e}$ is physically reasonable. For example, taking $T_e=30$ Kelvin, $\hbar v_F\frac{1}{l_B}=1200$ Kelvin (this magnetic energy scale can be reached with only $9$ Telsa for $v_F=10^6$ m/s), $\frac{v_F}{c_s}=100$, and $T_D=140$ Kelvin, we find $\frac{1}{\alpha_B^2} \times {1 \over 2} \frac{v_F}{ c_s}\frac{T_D}{T_e}\approx .15$. For Weyl and Dirac semimetals, the Debye temperature can range from 140 Kelvin in $\mathrm{Au}_2$Pb \cite{2014arXiv1412.2767S}, 200 Kelvin in $\mathrm{Cd}_3\mathrm{As}_2$ \cite{wang2007computation} to 420 Kelvin in the pyrochlore iridates \cite{doi:10.1143/JPSJ.70.2880,doi:10.1143/JPSJ.71.2578}. While we have derived this result for zero lattice temperature, it is applicable when $T_e\gg T_L$. We note this result is only valid when the scattering between Dirac/Weyl nodes of different chirality is weak.

The contribution from the lowest Landau level will dominate as long as $\alpha_B\gg1$. All higher Landau levels are exponentially suppressed by $e^{-\alpha_B n}$, where $n$ is the $n$th Landau level. (We note at finite chemical potential, $\alpha_B n$ must be greater than $\mu\beta_e$ to suppress the $n$th Landau level.)
From this, we find (using the heat capacity, $C_e=\frac{V k_B^2T}{12l^2\hbar v_F}$ \cite{refId0}, which is valid when $\alpha_B\gg1$)
\begin{equation}
\frac{\partial T_e}{\partial t}=-\gamma_B\frac{1}{T_e},~~~~~~\gamma_B=\frac{3 D^2k_B^3 T_D^5}{80\pi^7\rho v_F (\hbar c_s)^4 }.
\end{equation}
Notably, this rate is independent of the magnetic field and the distance between Weyl nodes (provided scattering between nodes is negligible). We note there will be small corrections due to the small contribution of higher Landau levels. The electronic temperature decays as
\begin{equation}
T_e(t)=\sqrt{T^2_0-2\gamma_Bt}.
\end{equation}
Taking $\omega_D$ to be 140 Kelvin and $c_s=5\times10^3$ m/s, we find $\gamma_B=40\times10^{12}$ K/s. We do not use the speed of sound and Debye temperature of $\mathrm{Cd}_3\mathrm{As}_2$ for this calculation since that material has a larger ratio of Fermi velocity to speed of sound and  Debye temperature then other Dirac/Weyl materials (this makes the limit $\alpha_B^2\gg\frac{v_F}{c_s}\frac{T_D}{T_e}$ harder to reach). Furthermore, to see quantum limit transport in $\mathrm{Cd}_3\mathrm{As}_2$ one needs fields of $43$ Tesla \cite{2014arXiv1412.0330Z}. This is due to the large Fermi surface of $\mathrm{Cd}_3\mathrm{As}_2$. As a result of such high fields and low temperatures, one might expect electron-phonon coupling to be modified \cite{Strong_Mag_Phonons}.

Finally, we remark that these results are rather unique to Weyl or Dirac semimetals and one does not generically expect to see electronic cooling dominated by the lowest Landau level in normal metals. This is because the magnetic energy scale for Weyl or Dirac semimetals is much larger than that of normal metals \cite{refId0}. More explicitly, the magnetic energy scale for Weyl or Dirac semimetal with $v_F=10^6$ m/s is 1200 Kelvin for a 9 Telsa magnetic field. For the same applied field in a normal metal it is $12.6\times\frac{m_e}{m}$ Kelvin, where $m$ is the effective mass and $m_e$ is the electron mass. For most metals, $m_e/m$ is on the order of unity \cite{Ashcroft}.


\section{Conclusion} \label{sec:CON}
In this work, we have analytically studied the cooling power of acoustic phonons as a function of doping level, disorder, and externally applied magnetic fields.  Our main results are in Eqs.~(5),~(7),~(9) and (16), along with the corresponding decays for the electronic temperature, $T_e$, in each case. Importantly, we find disorder can effectively be used to control the cooling power in $\mathrm{Cd}_3\mathrm{As}_2$ and other closely related materials. We stress that we have ignored electronic cooling from optical phonons and that our results are only valid for some temperature below the temperature of the lowest optical branch. This temperature depends crucially on the chemical potential, electronic-optical phonon coupling strength, and electron and lattice temperatures. We leave these material specific details as an open question. In future work it would be interesting to study the effect of Fermi arcs and Kondo impurities \cite{2014arXiv1410.8532P} on electronic cooling, as well as interactions \cite{PhysRevLett.113.136402,PhysRevB.90.035126,doi:10.7566/JPSJ.83.094710,PhysRevB.90.075137,doi:10.7566/JPSJ.82.033702}. 

{\em Note Added} -- Just prior to completion of this work, we noticed experimental results on the cooling by phonons in $\mathrm{Cd}_3\mathrm{As}_2$ for temperatures far above the temperature of the lowest optical branch \cite{2015arXiv150207007W}. It was suggested in Ref.~\cite{2015arXiv150207007W}, that hot carriers and optical phonons equilibrate rapidly ($500\times10^{-12}$ s) followed by slower cooling ($10^{-12}$ s) through the emission of acoustic phonons by the decay of optical phonons or hot carriers.

{\bf \em Acknowledgments} -- R.L. thanks  A. H. Macdonald, C. Weber, D. Lorshbough, W. Witczak-Krempa, D. Dicus for useful discussions. We thank P. Laurell for collaboration on related work. R.L. was partially supported by National Science Foundation (NSF) Graduate Research Fellowship award number 2012115499. R.L. and G.A.F. were supported by ARO Grant No. W911NF-14-1-0579, NSF Grant No. DMR-0955778, and DARPA grant No. D13AP00052.

\onecolumngrid

\appendix
\section{Power Loss of Single Weyl Node}\label{sec:Weyl_Node}

In this section, we provide some key steps in the derivation of Eq.~\eqref{Gen_Power}, starting from Eq.~\eqref{powerloss_NOB}. It is first helpful to seperate the power into two terms, $P_{ind}$ and $P_{spon}$, depending if they describe induced transitions or spontaneous transitions \cite{PhysRevB.81.245404}. These two terms are given by (with $\hbar=1$)
\begin{equation}
P_{ind}(\mu,T_e,T_L)=-2\pi\sum_{\vec{q}}\sum_{\vec{p}\beta}\sum_{\vec{k}\alpha}\epsilon_{\vec{k},\vec{p}}^{\alpha,\beta}w^{\alpha\beta}_q[f(\epsilon_{\vec{k}}^\alpha)-f(\epsilon_{\vec{p}}^\beta)] N^L(\omega_q)\delta_{\vec{k},\vec{p}+\vec{q}}\delta(\epsilon_{\vec{k},\vec{p}}^{\alpha,\beta}-\omega_q),
\label{P_ind_NO_APPROX}
\end{equation}
and 
\begin{align}
P_{spon}(\mu,T_e)=+2\pi \sum_{\vec{q}}\sum_{\vec{p}\beta}\sum_{\vec{k}\alpha}\epsilon_{\vec{k},\vec{p}}^{\alpha,\beta}w^{\alpha\beta}_q[f(\epsilon_{\vec{k}}^\alpha)-f(\epsilon_{\vec{p}}^\beta)]N^e(\omega_q)\delta_{\vec{k},\vec{p}+\vec{q}}\delta(\epsilon_{\vec{k},\vec{p}}^{\alpha,\beta}-\omega_q).
\end{align}
We note that $P_{spon}(\mu,T_e,T_e)=P_{ind}(\mu,T_e)$. As such, we only need to evaluate $P_{ind}$ \cite{PhysRevB.81.245404}. We now consider the limit $c_s\ll v_F$ as discussed in the main text. In this limit, we can neglect inter-band transitions, i.e. $\alpha\neq\beta$ \cite{PhysRevLett.102.206410}. It is instructive to consider each term in the $\alpha$ sum separately. For $\alpha=1$, using the delta function and the identity, $\int_{0}^{\infty}\mathrm{d}\epsilon \delta(\epsilon-\epsilon^{\alpha=1}_{\vec{p}+\vec{q}})g(\epsilon)=g(\epsilon^{\alpha=1}_{\vec{p}+\vec{q}})$, we have
\begin{align}
P^{\alpha=1}_{ind}(\mu,T_e,T_L)=-2\pi\sum_{\vec{q}}\sum_{\vec{p}}\int_{0}^{\infty} \mathrm{d}\epsilon (\epsilon-\epsilon_{\vec{p}}^{1})w^{11}_q[f(\epsilon)-f(\epsilon_{\vec{p}}^1)]N^L(\omega_q)\delta(\epsilon-\epsilon_{\vec{p}}^{1}-\omega_q)\delta(\epsilon-\epsilon_{\vec{p}+\vec{q}}^{1}).
\label{P_induced}
\end{align}
We can rewrite $\delta(\epsilon-v_Fp-\omega_q)$ as $\frac{1}{v_F}\delta\bigg(p-\big(\frac{\epsilon-cq}{v_F}\big)\bigg)$ to evaulate the $p$ integral. This gives
\begin{align}
P^{\alpha=1}_{ind}(\mu,T_e,T_L)=-2V\pi\frac{D^2}{4\rho v_F^3}\frac{4\pi(2\pi)}{(2\pi)^6}\int\mathrm{d}q\int_{cq}^{\infty}\mathrm{d}\epsilon \int_{-1}^{1}\mathrm{d}x (\epsilon-cq)^2q^4\bigg(1+\frac{(\epsilon-cq)+v_Fqx}{\sqrt{(\epsilon-cq)^2+v_F^2q^2+2(\epsilon-cq)v_Fqx}}\bigg)\nonumber \\
\times[f(\epsilon)-f(\epsilon-cq)]N^L(cq)\delta\bigg(\epsilon-\sqrt{(\epsilon-cq)^2+v_F^2q^2+2(\epsilon-cq)v_Fqx}\bigg).
\end{align}
After making the $q$ integral dimensionless, we have (to lowest order in $\frac{c}{v_F}$)
\begin{align}
P^{\alpha=1}_{ind}(\mu,T_e,T_L)\approx-2V\pi\frac{D^2}{4\rho v_F^8
\beta_L}\frac{4\pi(2\pi)}{(2\pi)^6}\int_0^{\infty}\mathrm{d}y\int_{0}^{\infty}\mathrm{d}\epsilon \int_{-1}^{1}\mathrm{d}x \epsilon^2y^3\bigg(1+\frac{\epsilon+yx}{\sqrt{\epsilon^2+y^2+2\epsilon yx}}\bigg)\frac{\partial f}{\partial \epsilon}\delta\bigg(\epsilon-\sqrt{\epsilon^2+y^2+2\epsilon yx}\bigg).
\end{align}
The remaining delta function can be rewritten as $\delta\left(\epsilon- \sqrt{\epsilon^2+y^2+2\epsilon yx}\right)=\delta(x+\frac{ y}{2\epsilon})\frac{1}{y}.$ Using the delta function to evalute the $x$ integral, we find
\begin{align}
P^{\alpha=1}_{ind}(\mu,T_e,T_L)=-2V\pi\frac{ D^2}{4\rho v_F^8\beta_L}\frac{4\pi(2\pi)}{(2\pi)^6}\int_0^{2\epsilon}\mathrm{d}y\int_{0}^{\infty}\mathrm{d}\epsilon\epsilon^2y^3\bigg(2-\frac{y^2}{2\epsilon^2}\bigg)\frac{\partial f}{\partial \epsilon}=-\frac{ D^2}{\rho v_F^8\beta_L(2\pi)^3}\frac{8}{6}\int_{0}^{\infty}\mathrm{d}\epsilon\epsilon^6\frac{\partial f}{\partial \epsilon}.
\end{align}
After integrating by parts and making the $\epsilon$ integral dimenionless, we have
\begin{align}
P^{\alpha=1}_{ind}(\mu,T_e,T_L)=\frac{ VD^2}{\rho v_F^8\pi^3}(k_BT_e)^6(k_BT_L)\int_{0}^{\infty}\mathrm{d}xx^6f(x-\beta_e\mu).
\end{align}
After performing a similar calculating for $P^{\alpha=-1}_{ind}(\mu,T_e,T_L)$, we find $P_{ind}$ is given by
\begin{align}
P_{ind}(\mu,T_e,T_L)=\frac{ VD^2}{\rho v_F^8\pi^3}(k_BT_e)^6(k_BT_L)\int_{0}^{\infty}\mathrm{d}xx^6\bigg(f(x-\beta_e\mu)+f(x+\beta_e\mu)\bigg).
\end{align}
The total power loss is then
\begin{align}
P=-\frac{V D^2}{\rho v_F^8\pi^3}(k_BT_e)^6(k_BT_e-k_BT_L)\int_{0}^{\infty}\mathrm{d}xx^6\bigg(f(x-\beta_e\mu)+f(x+\beta_e\mu)\bigg).
\end{align}
\section{Derivation of Transition Matrix Element}\label{sec:DIS_APP}

In this section we derive the transition matrix element in the case of weak zero-range disorder described by the following potential
\begin{equation}
V(\vec{r})=\frac{u}{2}\sum_{r_j}\delta(r-r_j)(1+\sigma_z).
\end{equation}
This derivation generalizes the one in Ref.~\cite{PhysRevLett.109.106602} for graphene to three dimensional topological semimetals. We assume that the concentration of disorder is low and can dress the electron-phonon vertex by scattering off a single impurity. The exact transition matrix element is given by
\begin{equation}
M_{\pm}=\langle k'|M^0_{\pm}G(p)\hat{T}+\hat{T}G(p)M_{\pm}^0+\hat{T}GM^{0}_\pm G(p)\hat{T}|k\rangle
\label{exact_M}
\end{equation}
where $G(p)$ is the free electron Green's function, $\hat{T}$ is the scattering operator (or $\hat{T}$-matrix) for a single impurity. The scattering operator to lowest order in disorder strength is taken to be the Fourier transformed impurity potential. We now make some approximations of the free electron Green's functions, similar to the ones made in Ref.~\cite{PhysRevLett.109.106602} for graphene. As mentioned in the main text, this formalism has been successful in providing understanding experimental results of electron cooling in graphene. We expect phonons with momentum $k_B T/c_s$, to dominate cooling. As such, we expect the virtual electrons to have much larger momentum than incoming and outing going electrons ($k,k' \ll p$). This allows one to approximate the electron Green's function, when the virtual states have an energy $\hbar v_F p \gg k_B T, \mu$, as $G(p)=-\frac{1}{\hbar v_F p}$. Plugging this into Eq.~\eqref{exact_M}, we find
\begin{equation}
|M_{\pm}|^2=|M|^2=\frac{\pi u^2 D^2n_0}{4\rho \hbar c_s v_F^2 q^3}|\langle k'|(\vec{\sigma}\times \vec{q})_z|k\rangle|^2,
\end{equation}
where $n_0$ is the impurity concentration. Here, the summing over impurities is done after squaring $M_{\pm}$ \cite{2014arXiv1410.5426S}. For simplicity, we use the angular average of $|\langle k'|(\vec{\sigma}\times \vec{q})_z|k\rangle|^2$ which is $q^2/2$. If $\mu \gg k_BT$, we can approximate the sum over $k$ and $k'$ as $\nu(\mu)^2\int\int \mathrm{d}\epsilon\mathrm{d}\epsilon'$. The power loss is then
\begin{equation}
P=\nu(\mu)^2u^2\sum_{\vec{q}}\left(|M_+|^2\omega_q\int\mathrm{d}\epsilon f(\epsilon)(1-f(\epsilon+\omega_q))N^{ph}_q+|M_-|^2(-\omega_q)\int\mathrm{d}\epsilon f(\epsilon)(1-f(\epsilon-\omega_q))(N^{ph}_q+1)\right).
\end{equation}
Evaluating the remaining $\epsilon$ integral, we have (defining the mean free path, $l=\frac{v_f}{2\pi u^2  n_0 \nu(\mu)}$),
\begin{equation}
P=\frac{V\pi\nu(\mu)D^2k_b^4}{\rho (\hbar c_s)^3v_F l}\frac{\pi^4}{30}(T_e^4-T_{L}^4).
\end{equation}
\section{Phonon Self-Energy in a Magnetic Field}\label{sec:MAG_APP}
In this section we derive the imaginary part  of the phonon self-energy in a magnetic field. The Green's function (for a given chirality, $\chi$) for a Weyl semimetal described the Hamiltonian in the main text is given by \cite{PhysRevB.89.085126}
\begin{align}
G^{\chi}(\omega,\vec{k},\vec{k}_{\perp})=i e^{-k^2_{\perp}l_B^2}\sum_{\lambda={\pm}}\sum_{n=0}^{\infty}\frac{(-1)^n}{E_{n}^{\chi}}((E_n^{\chi}\gamma_0-\lambda v_F (k_z-\chi b)\gamma^3)\{\mathcal{P}_{-}L_{n}(2k_{\perp}^2l_B^2)-\mathcal{P}_{+} L_{n-1}(2k_{\perp}^2l_B^2)\}+\nonumber \\
2\lambda v_F (\vec{k}_{\perp}\cdot\vec{\gamma}_{\perp}) L^1_{n-1}(2k_{\perp}^2l_B^2))\frac{1}{\omega+\mu-\lambda E_n^{\chi}},
\end{align}
where $L_n^\alpha$ are the generalized Laguerre polynomials, $\mathcal{P_{\pm}}=\frac{1}{2}\bigg(1\pm i \mathrm{sign}(eB) \gamma^1\gamma^2\bigg)$ and 
\begin{equation}
E_n^{\chi}=v_F\sqrt{(k_z-\chi b)^2+2n\frac{|e B|}{c}}.
\end{equation}
Following Ref.~\cite{PhysRevD.83.085003}, we rewrite our Green's function in a mix of real-space and momentum space coordinates. The partial Fourier transform of the Green's function is given by
\begin{equation}
G(\omega,k_z,\vec{r}_\perp)=V_{\perp}\int \frac{\mathrm{d}\vec{k}_{\perp}}{(2\pi)^2}e^{i\vec{k}_\perp\cdot\vec{r}_\perp}G^{\chi}(\omega,k_z,\vec{k}_{\perp}).
\end{equation}
The inverse partial Fourier transform is
\begin{equation}
G(\omega,k_z,\vec{k}_\perp)=\frac{1}{V_{\perp}}\int \mathrm{d}\vec{r}_{\perp}e^{-i\vec{k}_\perp\cdot\vec{r}_\perp}G(\omega,k_z,\vec{r}_\perp).
\label{IFT}
\end{equation}
The hybrid real-space/momentum-space Green's function is then
\begin{align}
G^{\chi}(\omega,\vec{k},\vec{r}_{\perp})=i \frac{V_{\perp}}{2\pi}\frac{1}{4l_B^2}e^{-\frac{r^2_{\perp}}{4l_B^2}}\sum_{\lambda={\pm}}\sum_{n=0}^{\infty}\frac{(-1)^n}{E_{n}^{\chi}}\Bigg((E_n^{\chi}\gamma_0-\lambda v_F (k_z-\chi b)\gamma^3)\{\mathcal{P}_{-}L_{n}\bigg(\frac{r_{\perp}^2}{2l_B^2}\bigg)-\mathcal{P}_{+} L_{n-1}\bigg(\frac{r_{\perp}^2}{2l_B^2}\bigg)\}-\nonumber \\2i\frac{v_F}{l_B^2}\lambda  (\vec{r}_{\perp}\cdot\vec{\gamma}_{\perp}) L^1_{n-1}\bigg(\frac{r_{\perp}^2}{2l_B^2}\bigg)\Bigg)\frac{1}{\omega+\mu-\lambda E_n^{\chi}}.
\end{align}
The total Green's function for both chiralities is then
\begin{equation}
G(\omega,k_z,\vec{r}_\perp)=\sum_{\chi=\pm}G^{\chi}(\omega,k_z,\vec{r}_\perp)\mathcal{P}^{\chi}_5.
\end{equation}
It is convenient to introduce the spectral function
\begin{equation}
A(\omega,k_z,\vec{r}_\perp)=\frac{1}{2\pi i}\left(G_{\mu=0}(\omega-i\epsilon,k_z,\vec{r}_\perp)-G_{\mu=0}(\omega+i\epsilon,k_z,\vec{r}_\perp)\right)=\sum_{\chi=\pm}A^{\chi}(\omega,k_z,\vec{r}_\perp)\mathcal{P}^{\chi}_5,
\end{equation}
where
\begin{align}
A^{\chi}(\omega,k_z,\vec{r}_\perp)=i \frac{V_{\perp}}{2\pi}\frac{1}{4l_B^2}e^{-\frac{r^2_{\perp}}{4l_B^2}}\sum_{\lambda={\pm}}\sum_{n=0}^{\infty}\frac{(-1)^n}{E_{n}^{\chi}}\Bigg((E_n^{\chi}\gamma_0-\lambda v_F (k_z-\chi b)\gamma^3)\{\mathcal{P}_{-}L_{n}\bigg(\frac{r_{\perp}^2}{2l_B^2}\bigg)-\mathcal{P}_{+} L_{n-1}\bigg(\frac{r_{\perp}^2}{2l_B^2}\bigg)\}-\nonumber \\2i\frac{v_F}{l_B^2}\lambda  (\vec{r}_{\perp}\cdot\vec{\gamma}_{\perp}) L^1_{n-1}\bigg(\frac{r_{\perp}^2}{2l_B^2}\bigg)\Bigg)\delta(\omega-\lambda E_n^{\chi}),
\end{align}
as done in Ref.~\cite{PhysRevB.89.085126}. The spectral function and Green's function are related by
\begin{equation}
G(i\omega_n,k_z,\vec{r}_\perp)=\int^{\infty}_{-\infty}\frac{\mathrm{d}\omega A(\omega,k_z,\vec{r}_\perp)}{i\omega_n+\mu-\omega}.
\label{spectral_Green}
\end{equation}
As discussed in the main text, the phonon self-energy to one-loop order is given by
\begin{equation}
\Pi^{\mathrm{Ph}}(B,\vec{q},i\omega_n)=\frac{g^2(\vec{q})}{\beta V}\sum_{ip_m}\sum_{\vec{k}}\mathrm{Tr}[\gamma^0G(\vec{k},ip_m)\gamma^0G(\vec{k}+\vec{q},i p_m+i\omega_n)],
\end{equation}
where the trace is over spinor indicies. After performing the Matsubara sum and analytically continuing ($i\omega_n\rightarrow \Omega+i\eta$)
\begin{equation}
\Pi^{\mathrm{Ph}}(B,\vec{q},\Omega)=\frac{g^2(\vec{q})}{V}\int\mathrm{d}\omega\int\mathrm{d}\omega'\frac{n_e(\omega-\mu)-n_e(\omega'-\mu)}{\omega-\omega'-\Omega-i\eta}\sum_{\vec{k}}\mathrm{Tr}[\gamma^0A(\vec{k},\omega)\gamma^0A(\vec{k}+\vec{q},\omega')].
\end{equation}
We are only concerned with the imaginary part of the phonon self-energy. Using the identity $\frac{1}{a+i\eta}=\mathrm{P}(\frac{1}{a})+i\pi\delta(a)$ (when $a$ is real), we have
\begin{equation}
\mathrm{Im}\Pi^{\mathrm{Ph}}(B,\vec{q},\Omega)=g^2(\vec{q})\int\mathrm{d}\omega \bigg(n_e(\omega-\mu)-n_e(\omega-\Omega-\mu)\bigg)\int\frac{\mathrm{d}k_z}{2\pi}\int\frac{\mathrm{d}^2k_{\perp}}{(2\pi)^2}\mathrm{Tr}[\gamma^0A(\vec{k},\omega)\gamma^0A(\vec{k}+\vec{q},\omega-\Omega)].
\end{equation}
Switching to the real-space/momentum space spectral function via Eq.~\ref{IFT}, we have
\begin{equation}
\mathrm{Im}\Pi^{\mathrm{Ph}}(B,\vec{q},\Omega)=\frac{g^2(\vec{q})}{V_{\perp}^2}\int\mathrm{d}\omega \bigg(n_e(\omega-\mu)-n_e(\omega-\Omega-\mu)\bigg)\int\frac{\mathrm{d}k_z}{2\pi}\int\mathrm{d}^2r_{\perp}\mathrm{Tr}[\gamma^0A(k_z,\omega,\vec{r}_\perp)\gamma^0A(k_z+q_z,\omega-\Omega,-\vec{r}_\perp)]e^{-i\vec{q}_\perp\cdot\vec{r}_\perp}.
\end{equation}
There are two different real-space integrals that need to be evaluated. They are as follows
\begin{align}
\int_0^\infty \mathrm{d}^2re^{i\vec{q}_\perp\cdot\vec{r}_\perp}L_{n}\bigg(\frac{r^2}{2l_B^2}\bigg)L_{n'}\bigg(\frac{r^2}{2l_B^2}\bigg)e^{-\frac{r^2}{2l_B^2}}=2\pi l_B^2(-1)^{(n+n')}e^{-\frac{q_{\perp}^2l_B^2}{2}}L_{n}^{n'-n}\bigg(\frac{q_{\perp}^2l_B^2}{2}\bigg)L_{n'}^{n-n'}\bigg(\frac{q_{\perp}^2l_B^2}{2}\bigg),
\end{align}
and
\begin{align}
 \int\mathrm{d}^2r e^{i\vec{q}_\perp\cdot\vec{r}_\perp}\frac{r^2_{\perp}}{2l_B^2}L^1_{n-1}\bigg(\frac{r^2}{2l_B^2}\bigg)L^1_{n'-1}\bigg(\frac{r^2}{2l_B^2}\bigg)e^{-\frac{r^2}{2l_B^2}}=2\pi l^2 n' (-1)^{(n+n')} e^{-\frac{q_{\perp}^2l_B^2}{2}}L^{n'-n}_{n-1}\bigg(\frac{q_{\perp}^2l_B^2}{2}\bigg)L_{n'}^{n-n'}\bigg(\frac{q_{\perp}^2l_B^2}{2}\bigg).
\end{align}
After performing the trace, we are left with three terms that group by their Laguerre polynomials and the imaginary phonon self-energy can be written as sum of two terms, $I_1+I_2$. The first term is
\begin{align}
I_1=\frac{g^2(\vec{q})}{4\pi^2l_B^2}\frac{1}{8\pi^2}\sum_{n,n'}\sum_{\chi,\lambda,\lambda'}\int_{-\infty}^\infty\frac{\mathrm{d}k_z}{2\pi}\int\mathrm{d}\omega\frac{\sinh(\beta\frac{\Omega}{2})}{\cosh(\beta\frac{\Omega}{2})+\cosh(\beta(\omega-\mu-\frac{\Omega}{2}))}\frac{1}{E_n^{\chi}(k_z)E_{n'}^{\chi}(k_z+q_z)}\nonumber \\
\times\Bigg([E_n^{\chi}(k_z)E_{n'}^{\chi}(k_z+q_z)+\lambda\lambda'v_F^2(k_z-\chi b)(k_z+q_z-\chi b)]+sv_F\chi[\lambda'E_n^{\chi}(k_z)(k_z+q_z-\chi b)+\lambda(k_z-\chi b)E_{n'}^{\chi}(k_z+q_z)]\Bigg) \nonumber \\
\times\delta(\omega-\Omega-\lambda' E_{n'}^{\chi}(k_z+q_z))\delta(\omega-\lambda E_n^{\chi}(k_z))  e^{-\frac{q_{\perp}^2l_B^2}{2}}(L_{n'}^{n-n'}(\frac{q_{\perp}^2l_B^2}{2})L_{n}^{n'-n}(\frac{q_{\perp}^2l_B^2}{2})+L_{n'-1}^{n-n'}(\frac{q_{\perp}^2l_B^2}{2})L_{n-1}^{n'-n}(\frac{q_{\perp}^2l_B^2}{2})).
\label{LLLCONT}
\end{align}
The second term is
\begin{align}
I_2=\frac{g^2(\vec{q})}{4\pi^2l_B^2}\frac{1}{\pi^2}\frac{v_F^2}{l_B^2}\sum_{n,n'}\sum_{\chi,\lambda,\lambda'}\int_{-\infty}^\infty\frac{\mathrm{d}k_z}{2\pi}\int\mathrm{d}\omega\frac{\sinh(\beta\frac{\Omega}{2})}{\cosh(\beta\frac{\Omega}{2})+\cosh(\beta(\omega-\mu-\frac{\Omega}{2}))}\frac{1}{E^{\chi}_n(k_z)E_n^{\chi}(k_z+q_z)}\nonumber \\
\times \delta\big(\omega-\Omega-\lambda' E_{n'}^{\chi}(k_z+q_z)\big)\delta(\omega-\lambda E_n^{\chi}(k_z))\lambda\lambda' n' e^{-\frac{q_{\perp}^2l_B^2}{2}}L_{n-1}^{n'-n}\bigg(\frac{q_{\perp}^2l_B^2}{2}\bigg)L_{n'}^{n-n'}\bigg(\frac{q_{\perp}^2l_B^2}{2}\bigg).
\end{align}
After shifting the $k_z$ in the integral by $b\chi$, doing the $\omega$ integral and summing over chirality, we find,
\begin{align}
I_1=2\frac{g^2(\vec{q})}{4\pi^2l_B^2}\frac{1}{8\pi^2}\sum_{n,n'}\sum_{\chi,\lambda,\lambda'}\int_{-\infty}^\infty\frac{\mathrm{d}k_z}{2\pi}\frac{\sinh(\beta\frac{\Omega}{2})}{\cosh(\beta\frac{\Omega}{2})+\cosh(\beta(\lambda E_n(k_z)-\mu-\frac{\Omega}{2}))}\frac{1}{E_n(k_z)E_{n'}(k_z+q_z)} \nonumber \\
\times\delta(\lambda E_n(k_z)-\Omega-\lambda' E_{n'}(k_z+q_z))\Bigg((E_n(k_z)E_{n'}(k_z+q_z)+\lambda\lambda'v_F^2(k_z)(k_z+q_z)\Bigg) e^{-\frac{q_{\perp}^2l_B^2}{2}}\nonumber \\
\times\Bigg(L_{n'}^{n-n'}\bigg(\frac{q_{\perp}^2l_B^2}{2}\bigg)L_{n}^{n'-n}\bigg(\frac{q_{\perp}^2l^2}{2}\bigg)+L_{n'-1}^{n-n'}\bigg(\frac{q_{\perp}^2l_B^2}{2}\bigg)L_{n-1}^{n'-n}\bigg(\frac{q_{\perp}^2l_B^2}{2}\bigg)\Bigg),
\label{LLLCONT}
\end{align}
and
\begin{align}
I_2=2\frac{g^2(\vec{q})}{4\pi^2l_B^2}\frac{1}{\pi^2}\frac{v_F^2}{l^2}\sum_{n,n'}\sum_{\lambda,\lambda'}\int_{-\infty}^\infty\frac{\mathrm{d}k_z}{2\pi}\frac{\sinh(\beta\frac{\Omega}{2})}{\cosh(\beta\frac{\Omega}{2})+\cosh(\beta(\lambda E_n(k_z)-\mu-\frac{\Omega}{2}))}\frac{1}{E_n(k_z)E_n(k_z+q_z)}\nonumber \\
\times \delta(\lambda E_n(k_z)-\Omega-\lambda' E_{n'}(k_z+q_z))\lambda\lambda' n' e^{-\frac{q_{\perp}^2l_B^2}{2}}L_{n-1}^{n'-n}\bigg(\frac{q_{\perp}^2l_B^2}{2}\bigg)L_{n'}^{n-n'}\bigg(\frac{q_{\perp}^2l_B^2}{2}\bigg).
\end{align}
We observe that due to the hyperbolic functions, the cooling power will be suppressed exponentially in terms of $\alpha n$. We thus focus on the contribution of the lowest Landau level to the cooling power. The only term to contribute from the lowest Landau level is $I_1$ due to the vanishing of the Laguerre polynomials ($L_{n}(x)$ with $n<0$ is defined to be zero \cite{PhysRevB.89.085126}). For $n=0$, the first term becomes,
\begin{align}
I_1=2\frac{g^2(\vec{q})}{4\pi^2l_B^2}\frac{1}{8\pi^2}2\sum_{\lambda}\int_{-\infty}^\infty\frac{\mathrm{d}k_z}{2\pi}\frac{\sinh(\beta\frac{\Omega}{2})}{\cosh(\beta\frac{\Omega}{2})+\cosh(\beta(\lambda v_F k_z-\mu-\frac{\Omega}{2}))}\delta(-\Omega-\lambda v_Fq_z)e^{-\frac{q_{\perp}^2l_B^2}{2}}.
\label{LLLCONT}
\end{align}
After evaluating the $k_z$ integral, we arrive at our final expression for the imaginary part of the phonon self-energy,
\begin{align}
\mathrm{Im}[\Pi^{\mathrm{Ph}}(\vec{q},\Omega)]=\sum_{\lambda=\pm}\frac{g^2(\vec{q})\Omega}{8\pi^5v_Fl_B^2}\delta(-\Omega-\lambda v_Fq_z)e^{-\frac{q_{\perp}^2l_B^2}{2}}.
\label{IM_PART_SE}
\end{align}
Plugging Eq.~\eqref{IM_PART_SE} into Eq.~\eqref{KOGAN} (at zero lattice temperature), we have
\begin{equation}
P_B=-\sum_{s,\lambda=\pm}\sum_{\vec{q}}\int\mathrm{d}\omega\omega^2 N^e(\omega)\frac{g^2(\vec{q})}{8\pi^5v_Fl_B^2}\delta(-\omega-\lambda v_Fq_z)e^{-\frac{q_{\perp}^2l_B^2}{2}}s\delta(\omega-sc_sq),
\label{KOGAN}
\end{equation}
Using one of the delta functions to evaluate the $\omega$ integral, we find
\begin{equation}
P_B=-\sum_{s,\lambda=\pm}V\int_0^{\Lambda}\frac{\mathrm{d}q}{(2\pi)^3}\int_0^{\pi}\mathrm{d}\theta \sin\theta \int_0^{2\pi}\mathrm{d}\phi c^2q^4 N^e(scq)\frac{g^2(\vec{q})}{8\pi^5v_Fl_B^2}\delta(-sc_sq-\lambda v_Fq\cos\theta)e^{-\frac{q^2\sin\theta^2l_B^2}{2}}s,
\label{KOGAN}
\end{equation}
Evaluating the $\phi$ integral, we have
\begin{equation}
P_B=-\sum_{s,\lambda=\pm}V\int_0^{\frac{\omega_D}{c}}\frac{\mathrm{d}q}{(2\pi)^2}\int_{-1}^{1}\mathrm{d}x  c_s^2q^3 N^e(sc_sq)\frac{g^2(\vec{q})}{16\pi^5v_F^2l_B^2}\delta(x+\lambda s\frac{c_s}{v_F})e^{-\frac{q^2(1-x^2)l_B^2}{2}}s.
\label{KOGAN}
\end{equation}
Here we have taken the cutoff, $\Lambda$, to be the Debye wavelength, $\frac{\omega_D}{c_s}$. Using the last delta function to evaluate the $x$ integral, we find
\begin{equation}
P_B=-\sum_{s,\lambda=\pm}\frac{scVD^2}{8\pi^5v_F^7l_B^24\rho \beta^5}\int_0^{\beta\frac{v_F\omega_D}{c}}\frac{\mathrm{d}y}{(2\pi)^2} y^4 \frac{1}{e^{s\frac{c}{v_F}y}-1}e^{-\frac{y^2(1-(\frac{c}{v_F})^2)l_B^2}{2\beta^2v_F^2}},
\label{KOGAN}
\end{equation}
\begin{equation}
P_B=-\sum_{\lambda=\pm}\frac{cVD^2}{8\pi^5v_F^7l_B^24\rho \beta^5(2\pi)^2}(\frac{v_F}{c})^5\sum_{k=0}^{\infty}\frac{ (-\frac{l_B^2}{2v_F^2\beta^2}\frac{v_F}{c})^k}{k!}\int_0^{\frac{T_D}{T_e}}\mathrm{d}zz^{k+4}\bigg(\frac{2}{e^{z}-1}+1\bigg).
\label{KOGAN}
\end{equation}
Here we have used the fact that $N^e(-z)=-1-N^e(z)$ and assumed $\frac{c}{v_F}\ll1$. We can approximate the first term in the integral as
\begin{equation}
\int_0^{\frac{T_D}{T_e}}\mathrm{d}zz^{k+4}\frac{2}{e^{z}-1}\approx\int_0^{\infty}\mathrm{d}zz^{k+4}\bigg(\frac{2}{e^{z}-1}\bigg)=2\eta(k+5)\Gamma(k+5),
\end{equation}
as for large $z$, the integrand is small. We then have
\begin{equation}
P_B=-\sum_{\lambda=\pm}\frac{c VD^2}{8\pi^5v_F^7l_B^24\rho \beta^5(2\pi)^2}(\frac{v_F}{c})^5\sum_{k=0}^{\infty}\frac{ (-\frac{l_B^2}{2v_F^2\beta^2}\frac{v_F}{c})^k}{k!}\bigg(2\eta(k+5)\Gamma(k+5)+\frac{1}{(k+5)}(\frac{T_D}{T_e})^{k+5}\bigg).
\end{equation}
Assuming $\frac{T_e}{T_D}\ll1$ and performing the trivial summing over $\lambda$, we have
\begin{equation}
P_B=-\frac{cVD^2}{4\pi^5v_F^7l_B^24\rho \beta^5(2\pi)^2}(\frac{v_F}{c})^5(\frac{T_D}{T_e})^5\sum_{k=0}^{\infty}\frac{ (-\frac{l_B^2}{2v_F^2\beta^2}\frac{v_F}{c}\frac{T_D}{T_e})^k}{k!(k+5)}.
\end{equation}
We assume that $\frac{l_B^2}{2v_F^2\beta^2} \frac{v_F}{c}\frac{T_D}{T_e}\ll1$, which is a physically reasonable limit. For example, taking $T_e=30$ Kelvin, $\frac{l_B}{2\hbar v_F}=1200$ Kelvin (the magnetic energy scale), $\frac{v_F}{c}=100$, and $T_D=140$ Kelvin, we find $\frac{l_B^2}{2v_F^2\beta^2} \frac{v_F}{c}\frac{T_D}{T_e}\approx .15$. Using the fact that $\frac{l_B^2}{2v_F^2\beta^2} \frac{v_F}{c}\frac{T_D}{T_e}\ll1$, we only keep the first term in the sum. We then arrive at our expression for the power loss (after restoring factors of $\hbar$) in the presence of an external magnetic field,
\begin{equation}
P_B=-\frac{VD^2\omega_D^5}{320\pi^7v_F^2l_B^24\rho c^4}.
\end{equation}
\twocolumngrid
\bibliography{outline.bib}
\end{document}